\documentstyle[11pt,newpasp,twoside,epsf]{article}
\def\edcomment#1{\iffalse\marginpar{\raggedright\sl#1\/}\else\relax\fi}
\reversemarginpar

\newcommand{\Ho}{H_{\rm o}}
\newcommand{\Omb}{\Omega_{\rm b}}

\newcommand{\lamo}{\lambda_{\rm o}}

\newcommand{\OmT}{\Omega_{\rm tot}}

\begin{document}
\title{Cosmological Parameter Estimation from CMB and X-ray clusters}
 \author{Marian Douspis}
\affil{Astrophysics, NAPL, Keble Road, OXFORD OX1 3RH, UK}
\author{A. Blanchard, R. Sadat, J.G. Bartlett}
\affil{Laboratoire d'Astrophysique de Toulouse, OMP, 14 av E. Belin, 31400 Toulouse France}

\begin{abstract}
We present the results of a combined analysis of cosmic microwave background 
(CMB) and X-ray galaxy clusters baryon fraction  to deduce constraints over 6 inflationnary cosmological  parameters. Such a combination is necessary for breaking degeneracies inherent to the CMB.
\end{abstract}

\section{Introduction}
        Since their first detection by COBE, the CMB temperature fluctuations 
have become an essential tool for constraining cosmological parameters. 
 From the 
beginning of 2000, new experiments
 have released data set of good quality up to the third acoustic peak.
Better constraints have been obtained on several cosmological parameters.
Nevertheless, it has been shown that even with precise measurements of the 
power spectrum, it is nearly impossible to distinguish models with the same
physical parameters on the last scattering surface. Basically, some 
degeneracies are inherent to the CMB. We consider in this work
X--ray clusters as an independent way for constraining cosmological parameters.


\section{Cosmological constraints from CMB and from X-ray Clusters}

In the present analysis we consider the data from COBE, 
BOOMERanG, MAXIMA and DASI. 
 We analysed the likelihood of inflationary 
models with $\Ho,\OmT$, $\lamo,\Omb h^2,n,Q$ as free cosmological parameters. 
This corresponds to $7.10^6$ models tested.To derive the likelihood values 
for the models we considered, we used the 
method developed in Bartlett et al. (2000), and already used 
in Le Dour et al. (2000).

The results are presented as two-dimensional contours plots of the likelihood 
projected onto various parameters planes. 
This 2-D presentation has the advantage of showing clearly all the degeneracies
between parameters. 
The left contour plots of
 figure 1  show some degeneracies between our parameters.
  Basically,  $H_o$, $\Lambda$ and $\OmT$  are strongly 
 degenerated if only CMB data are used in parameter estimation.
This means that one should consider another source of constraints, independent 
from CMB observations, for a better cosmological parameter estimation. 
We choose X-ray clusters as additional constraints.

 Clusters are interesting object for which both the luminous baryonic mass (X-ray emitting intracluster gas) and the total gravitating mass can be determined. Therefore, an upper limit on the fraction baryon, $f_b$, can be estimated. Results are not necessary in agreement allowing values between $f_b \sim 10\%$  and  $f_b \sim 20\%$. 
In this work, we consider the constraints given in  Sadat and Blanchard 2001~\cite{sadat} for the baryon fraction: $f_b = \frac{\Omega_b}{\Omega_m}= 0.10 \times h_{50}^{-3/2} \;\;(\pm 10\%)$.  The aim of this contribution being 
to show that clusters may help for breaking degeneracies shown by CMB parameters estimations, we did not  consider constraints given by other groups. A better
consideration of these different results on X-ray clusters will be found in 
Douspis et al. (2001b). 

\section{Combined analysis and conclusion}

Given the constraints of CMB and X-ray clusters, one could combined the two by multiplying the corresponding likelihood.
\begin{figure}
\plottwo{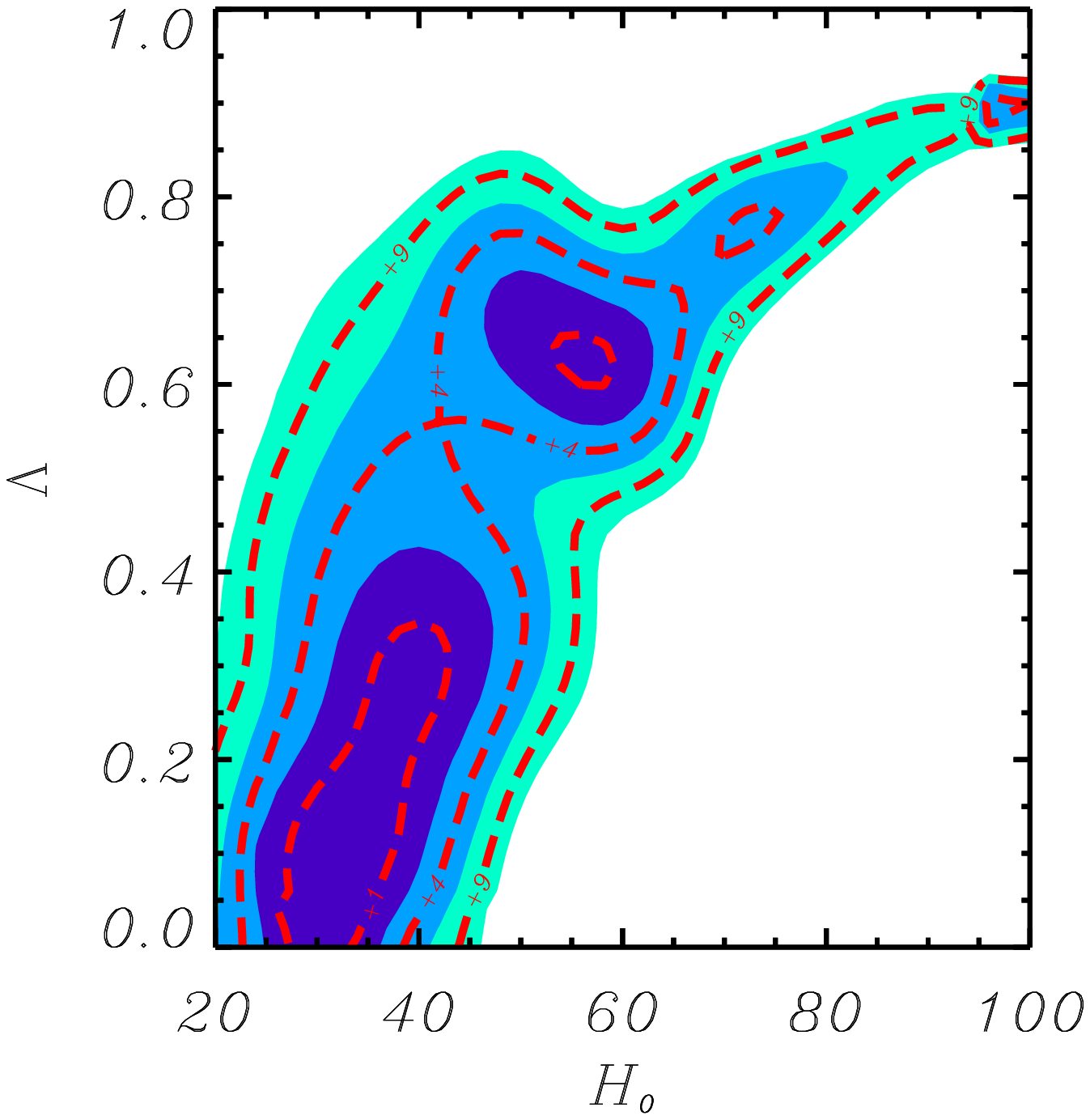}{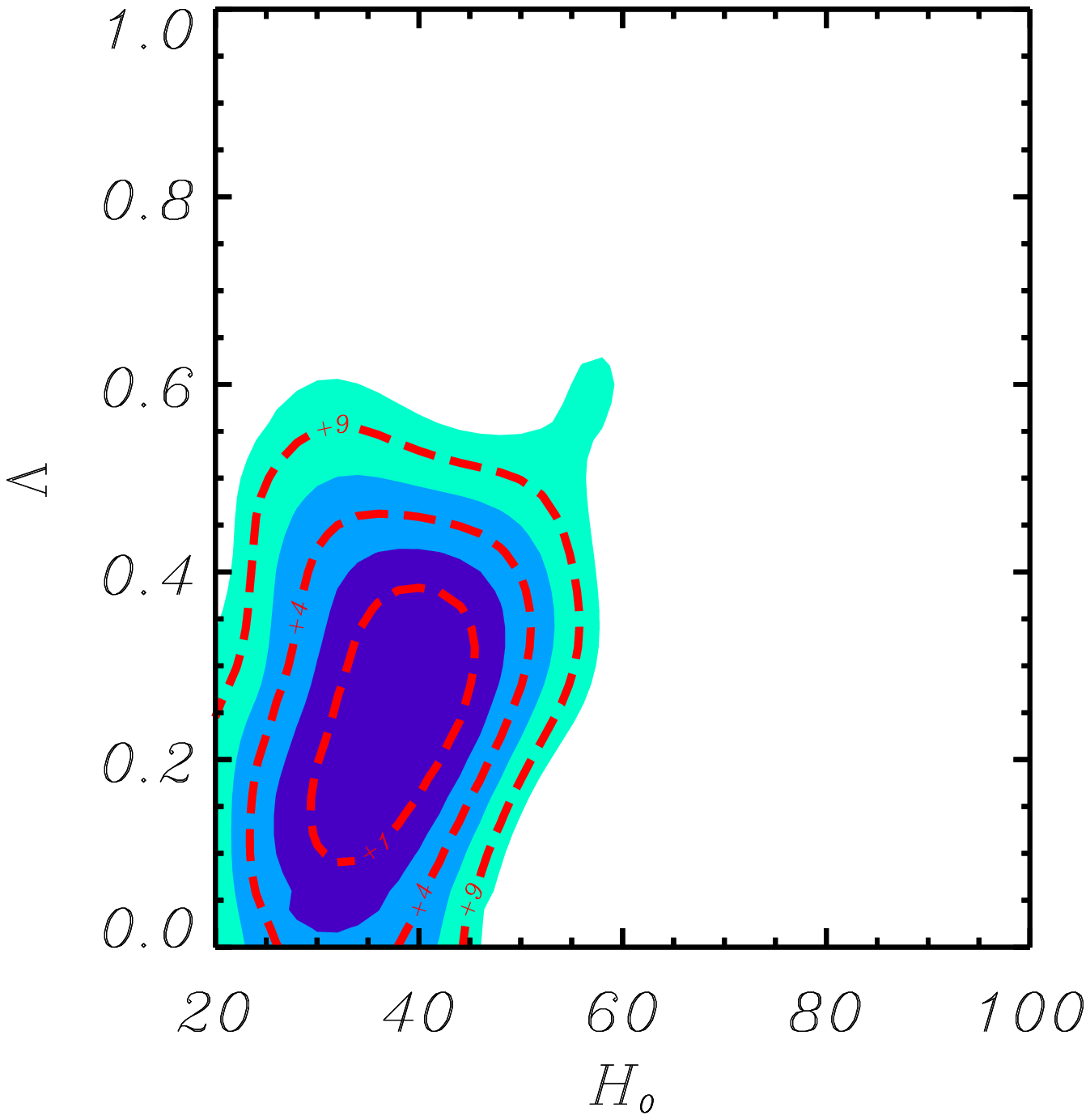}
\caption{Constraints given by CMB   and CMB plus  X--ray data
\label{fig:contcmb}}
\end{figure}
The right figure of fig. 1 shows the constraints from the 
combined analysis 
in the same plane as the left figure. 
We can see  that the degeneracies are broken. Confidence intervals are now
determined for $H_o$ and $\Lambda$. For both parameters, low values are
 preferred: $25 < H_o < 50 $ and $\Lambda < 0.45$ at 99\% CL.

Due to inherent degeneracies in the CMB it is nearly impossible to specify some of the cosmological parameter; ``cross constraints'' are then necessary. This work is thus a preliminary view
 of the kind of constraints one would obtain in a near future, using last 
X--ray satellite data, and in a less near future the CMB satellite data 
(MAP and Planck).

\end{document}